\documentclass[a4paper,fleqn]{cas-dc}

\usepackage[numbers,sort&compress]{natbib}

\usepackage{amsmath,amssymb}
\usepackage{booktabs}
\usepackage{graphicx}
\usepackage{algorithm}
\usepackage{algorithmic}
\usepackage{pgfplots}
\pgfplotsset{compat=1.18}

\graphicspath{{paper/}}

\newtheorem{definition}{Definition}
\newtheorem{assumption}{Assumption}
\newtheorem{problem}{Problem}
\newtheorem{lemma}{Lemma}
\newtheorem{proposition}{Proposition}
\newtheorem{theorem}{Theorem}
\newtheorem{corollary}{Corollary}
\newtheorem{remark}{Remark}

\newproof{proof}{Proof}
\let\origendproof\endproof
\renewcommand{\endproof}{\ifhmode\unskip\fi\hfill\ensuremath{\square}\origendproof}

\newcommand{\Acl}{A_{\mathrm{cl}}}
\newcommand{\Ccl}{C_{\mathrm{cl}}}
\newcommand{\D}{\mathcal D}

\newcommand{\Lyap}{\mathcal L}

\newcommand{\R}{\mathcal R}

\newcommand{\blkdiag}{\operatorname{blkdiag}}
\newcommand{\tr}{\operatorname{tr}}
\newcommand{\spec}{\operatorname{spec}}

\newcommand{\RRe}{\operatorname{Re}}

\begin{document}
\let\WriteBookmarks\relax
\def\floatpagepagefraction{1}
\def\textpagefraction{.001}

\shorttitle{Coupled-Improvement Policy Iteration in LQ Stochastic Differential Games}

\shortauthors{K. Handwerker et al.}

\title [mode = title]{Coupled-Improvement Policy Iteration in Linear-Quadratic Stochastic Differential Games with State- and Control-Dependent Noise}

\author[1]{Karl Handwerker}
\cormark[1]
\ead{karl.handwerker@kit.edu}

\author[1]{Lucas G\"unther}
\ead{lucas.guenther@kit.edu}

\author[1]{Felix Th\"ommes}
\ead{felix.thoemmes@kit.edu}

\author[1]{Balint Varga}
\ead{balint.varga2@kit.edu}

\author[1]{S\"oren Hohmann}
\ead{soeren.hohmann@kit.edu}

\affiliation[1]{organization={Institute of Control Systems (IRS), Karlsruhe Institute of Technology (KIT)},
            city={Karlsruhe},
            postcode={76131},
            country={Germany}}

\cortext[1]{Corresponding author}

\begin{abstract}
This paper introduces a novel policy iteration with coupled improvement (CI-PI) for stochastic differential games with state- and control-dependent noise.
By jointly solving the players' stationarity conditions, CI-PI resolves the coupling induced by control-dependent noise, which schemes that hold the strategies of the other players fixed neglect.
We compare CI-PI with iterated best response (IBR), a well-established approach in dynamic games and multi-agent learning, and derive closed-form Fr\'echet derivatives at Nash equilibria to characterize the local convergence of both schemes.
Our analysis reveals a regime change as the noise intensity increases, beyond the weak-coupling regime that typically underlies convergence guarantees.
In this high-noise regime, CI-PI is guaranteed to converge locally and approaches superlinear convergence, whereas IBR can diverge with three or more players.
Numerical examples confirm that this regime change occurs already at moderate noise levels, where IBR fails in most three-player games.
\end{abstract}

\begin{keywords}
Stochastic differential game \sep
Policy iteration \sep
Feedback Nash equilibrium \sep
Multiplicative noise \sep
Iterated best response 
\end{keywords}

\maketitle
\hypersetup{pdfauthor={Karl Handwerker, Lucas G\"unther, Felix Th\"ommes, Balint Varga, S\"oren Hohmann}, pdfsubject={}}

\section{Introduction}

Noncooperative stochastic differential games are mathematical models of the dynamic interaction between multiple decision makers acting on an uncertain system, each pursuing an individual objective \cite{basar_olsder_1998,hu_lauriere_2024}.
The stochastic formulation captures uncertainties inherent in applications such as financial markets \cite{SAITO2019111}, networked control \cite{10811852}, and human-machine interaction \cite{10731297,zhang_2026}.
In such applications, the uncertainty typically scales with the state and with the applied controls \cite{Todorov2005}.
A central solution concept for such games is the feedback Nash equilibrium, a profile of state-feedback strategies for which no player can improve its own objective by a unilateral deviation \cite{basar_olsder_1998,starr_ho_1969}.
In stochastic infinite-horizon linear-quadratic (LQ) settings, the equilibrium conditions reduce to coupled stochastic algebraic Riccati equations \cite{ZhuZhang2013}.
Closed-form solutions to these equations are available only in special cases, so that feedback Nash equilibria must in general be computed numerically \cite{basar_olsder_1998}.

A straightforward approach to computing a Nash equilibrium is iterated best response (IBR).
In IBR, each player repeatedly responds optimally to the strategies of the others from the previous iteration \cite{basar_1987}.
IBR underlies learning in games and fictitious-play methods for stochastic differential games \cite{HanHu2020}.
Policy iteration (PI) replaces the exact best response by a single improvement step and is the most widely used approach for computing feedback Nash equilibria in deterministic games \cite{nortmann_2024,chen_2026_tac,guan_2024,thoemmes_2026}.

Whether IBR and PI converge is a property of the coupled equilibrium conditions, which form a fixed-point problem without the monotonicity structure of the single-player case \cite{seqPI,hu_lauriere_2024}.
Since nonzero-sum games may possess several stabilizing equilibria \cite{engwerda_2005}, convergence is a local and equilibrium-dependent question, which is open in general \cite{hu_lauriere_2024}.
Existing convergence guarantees for IBR and PI rest on weak coupling between the players, enforced by small interaction terms, short horizons or strong discounting \cite{basar_li_1989,srikant_basar_1992}.

{\emergencystretch=1.5em
For standard PI in games with state- and control-dependent noise, including the deterministic case, the Fr\'echet derivative of the iteration map derived in \cite{seqPI} exposes the terms that govern local convergence.
This analysis identified the coupling of the players through their noise channels, the noise coupling, as a structural limitation of local convergence and showed that growing noise weakens local contraction.
When the players act through a common noisy channel, the noise injected by one player changes the incentives of all others.
Every update against fixed gains of the others, as in IBR and existing PI schemes, ignores this interaction.
Some applications, however, involve moderate or even strong noise, for example random parameter perturbations \cite[Sec.~1.9]{Damm2004}.
In these cases, the noise coupling is no longer weak, and these guarantees do not carry over.\par}

In this paper, we propose policy iteration with coupled improvement (CI-PI) to lift this structural limitation.
CI-PI solves the stationarity conditions of all players jointly and thereby resolves their coupling through the common noise source.
We compare CI-PI with IBR as the representative of the schemes that hold the others fixed, since its exact best responses rule out inexact improvement steps as the cause of its limitations.
Our main result is that the noise coupling no longer limits the local convergence of CI-PI.
Near the stabilizability limit, its local rate even tends to zero, whereas that of IBR approaches a limit set by the geometry of the noise channels, which can exceed one for three or more players.
Specifically, our contributions are:
\begin{enumerate}
\item A novel policy iteration with coupled improvement whose fixed points are exactly the stabilizing feedback Nash equilibria.
\item A local convergence analysis of CI-PI and IBR, based on closed-form Fr\'echet derivatives.
\item An analysis of Nash equilibria under growing noise. It identifies a regime change from cost-driven to noise-driven local dynamics beyond weak coupling and gives conditions under which CI-PI approaches superlinear convergence and IBR diverges.
\end{enumerate}

\noindent \textbf{Notation:}
$\mathbb S^n$ denotes the set of real symmetric $n\times n$ matrices with the inner product $\langle X,Y\rangle:=\operatorname{tr}(XY)$, $\mathbb S^n_+$ and $\mathbb S^n_{++}$ its positive semidefinite and positive definite elements, also written $X\succeq0$ and $X\succ0$, and $\lambda_{\min}(X)$ and $\lambda_{\max}(X)$ the smallest and largest eigenvalues of $X\in\mathbb S^n$.
For a square matrix or a linear map $X$, $\operatorname{spec}(X)$ denotes its spectrum and $\rho(X)$ its spectral radius.
$\|\cdot\|$ denotes the Euclidean norm, the induced matrix norm, and, for linear maps between spaces of matrices, the operator norm induced by the Frobenius norm $\|\cdot\|_F$, and $I_n$, or $I$, the $n\times n$ identity matrix.
$W$ is a one-dimensional standard Brownian motion on a filtered probability space $(\Omega,\mathcal F,\{\mathcal F_t\}_{t\ge0},\mathbb P)$ satisfying the usual conditions.
\section{Problem Formulation}\label{sec:problem_statement}

We consider an $N$-player stochastic differential game with players $i\in\mathcal I:=\{1,\dots,N\}$, controls $u_i(t)\in\mathbb R^{m_i}$, $m:=\sum_{i\in\mathcal I}m_i$, joint control $u:=(u_i)_{i\in\mathcal I}$, and state $x(t)\in\mathbb R^n$ governed by the stochastic differential equation (SDE)
\begin{align}\label{eq:sde-N}
dx(t)
&=
\Bigl(Ax(t)+\sum_{i\in\mathcal I} B_i u_i(t)\Bigr)\,dt\nonumber\\
&\quad+
\Bigl(Cx(t)+\sum_{i\in\mathcal I} D_i u_i(t)\Bigr)\,dW(t),
\end{align}
with deterministic initial state $x(0)=x_0\in\mathbb R^n$ and constant matrices $A,C\in\mathbb{R}^{n\times n}$ and $B_i,D_i\in\mathbb{R}^{n\times m_i}$.
Each player $i \in \mathcal I$ seeks to minimize the infinite-horizon quadratic cost
\begin{equation}
\label{eq:cost}
J_i(u;x_0)
=
\mathbb E\int_0^\infty
x(t)^\top Q_i x(t)
+
\sum_{j\in\mathcal I} u_j(t)^\top R_{ij}u_j(t)
\,dt,
\end{equation}
where $Q_i\in\mathbb{S}^n_{++}$, $R_{ii}\in\mathbb S^{m_i}_{++}$, and $R_{ij}\in\mathbb S^{m_j}_{+}$ for all $j\in\mathcal{I}\setminus\{i\}$.
We combine the matrices $D_i$ and the control weights $R_{ii}$ of all players in $D:=[D_1\ \cdots\ D_N]\in\mathbb R^{n\times m}$ and $R:=\blkdiag(R_{11},\dots,R_{NN})\in\mathbb S^m_{++}$.

The players use linear state feedback $u_i(t)=-K_i x(t)$ with gains $K_i\in\mathbb R^{m_i\times n}$, stacked into the gain profile $K:=[K_1^\top\ \cdots\ K_N^\top]^\top\in\mathbb R^{m\times n}$.
We write $K_{-i}:=(K_j)_{j\in\mathcal I\setminus\{i\}}$ for the gains of all players except player $i$, and $(\widetilde K_i,K_{-i})$ for the profile in which player $i$ uses $\widetilde K_i$ instead of $K_i$.
With the closed-loop drift and diffusion matrices
\begin{equation}
A_{\mathrm{cl}}(K)
:=
A-\sum_{i\in\mathcal I} B_iK_i,
\quad
C_{\mathrm{cl}}(K)
:=
C-\sum_{i\in\mathcal I} D_iK_i,
\end{equation}
the closed-loop dynamics read
\begin{equation}
\label{eq:closed-loop-SDE}
dx(t)
=
A_{\mathrm{cl}}(K)x(t)\,dt
+
C_{\mathrm{cl}}(K)x(t)\,dW(t).
\end{equation}

\begin{definition}[Mean-square stability]
We call the closed-loop system \eqref{eq:closed-loop-SDE} mean-square stable (MSS) \cite[Def.~1.5.1]{Damm2004} if
$\lim_{t\to\infty}\mathbb{E}\|x(t)\|^2=0$ for all $x_0\in\mathbb{R}^n$.
\end{definition}

We denote the set of stabilizing gain profiles by
\begin{equation}\label{eq:K}
\mathcal K
:=
\left\{
K=(K_i)_{i\in\mathcal I}:
\eqref{eq:closed-loop-SDE}
\text{ is MSS}
\right\}.
\end{equation}

\begin{assumption}[Mean-square stabilizability]\label{ass:MSS}
The game is mean-square stabilizable by linear state feedback, that is, $\mathcal K\neq\varnothing$.
\end{assumption}

For $K\in\mathcal K$, we write $J_i(K;x_0)$ for the cost \eqref{eq:cost} of player $i$ under $u_j=-K_jx$, $j\in\mathcal I$, whose running cost is $x^\top S_i(K)x$ with
\begin{equation}\label{eq:running_cost}
S_i(K):=Q_i+\sum_{j\in\mathcal I}K_j^\top R_{ij}K_j\in\mathbb S^n_{++}.
\end{equation}
With the generalized Lyapunov operator
\[
\Lyap_K(Z):=\Acl(K)^\top Z+Z\Acl(K)+\Ccl(K)^\top Z\Ccl(K)
\]
on $\mathbb S^n$, the cost is
\begin{equation}\label{eq:policy-eval}
\Lyap_K\bigl(P_i(K)\bigr)+S_i(K)=0
\end{equation}
\cite[Lemma~1]{seqPI}.
We call $P_i(K)$ the value matrix of player $i$, which is positive definite since $S_i(K)\succ0$ \cite[Lemma~1]{seqPI}, and collect the value matrices of all players in $P(K):=(P_1(K),\dots,P_N(K))$.

\begin{definition}[Stabilizing Nash equilibrium]\label{def:LFNE}
A gain profile $K^\star\in\mathcal K$ is called a
stabilizing linear feedback Nash equilibrium \cite{basar_olsder_1998} if,
for every player $i\in\mathcal I$ and every gain $\widetilde K_i\in\mathbb R^{m_i\times n}$ such that $(\widetilde K_i,K_{-i}^\star)\in\mathcal K$,
\begin{equation}
J_i(K^\star;x_0)
\le
J_i\bigl((\widetilde K_i,K_{-i}^\star);x_0\bigr)
\qquad
\forall x_0\in\mathbb R^n.
\end{equation}
\end{definition}

In the sequel, equilibrium refers to a stabilizing linear feedback Nash equilibrium.

\begin{proposition}[Equilibrium conditions \cite{Lucas}]\label{prop:equilibrium}
A gain profile $K^\star\in\mathcal K$ is an equilibrium if and only if, with $P_i^\star:=P_i(K^\star)$ from \eqref{eq:policy-eval} and $P^\star:=P(K^\star)$,
\begin{equation}\label{eq:stationarity}
\bigl(R_{ii}+D_i^\top P_i^\star D_i\bigr)K_i^\star
=
B_i^\top P_i^\star
+
D_i^\top P_i^\star
\Bigl(
C-\sum_{j\neq i} D_jK_j^\star
\Bigr)
\end{equation}
holds for all $i\in\mathcal I$.
\end{proposition}

Condition \eqref{eq:stationarity} is the stationarity condition of player $i$.
Together with \eqref{eq:policy-eval}, it forms the coupled stochastic algebraic Riccati equations of the game, which are necessary and sufficient for an equilibrium \cite{Lucas}.
Through the term $D_i^\top P_i^\star\sum_{j\neq i}D_jK_j^\star$, the control-dependent noise couples the stationarity conditions of the players.
We refer to this effect as noise coupling.

\begin{problem}\label{prob:cipi}
Let Assumption~\ref{ass:MSS} hold. Design a policy iteration scheme for \eqref{eq:sde-N}--\eqref{eq:cost} whose fixed points are exactly the equilibria and whose local convergence, in contrast to that of IBR, is not driven by the noise coupling.
\end{problem}

\section{Policy Iteration with Coupled Improvement}\label{sec:cipi}

This section addresses the first requirement of Problem~\ref{prob:cipi} by introducing a policy iteration with coupled improvement. It enforces the stationarity conditions of all players jointly and thereby accounts for the noise coupling.
CI-PI alternates between policy evaluation and policy improvement.
The policy evaluation step assesses a given stabilizing and in general suboptimal gain profile $K\in\mathcal K$ by computing the values $P(K)$, that is, by solving the $N$ independent generalized Lyapunov equations \eqref{eq:policy-eval}, one per player.
To derive the improvement step and to compare it with IBR, we first state the best-response problem of a single player in Section~\ref{sec:best_response}.
Its exact solution defines IBR, and its stationarity condition is the starting point of the coupled improvement step in Section~\ref{sec:coupled_improvement}.
Throughout this section, the superscript $k$ denotes the iteration index.

\subsection{Best-Response Problem}\label{sec:best_response}

We consider the exact best-response problem obtained by fixing the gains of all players except player $i$.
For $K\in\mathcal K$ and $i\in\mathcal I$, let
\begin{equation}
\label{eq:induced_mats}
A_{-i}:=A-\sum_{j\neq i}B_jK_j,\qquad
C_{-i}:=C-\sum_{j\neq i}D_jK_j .
\end{equation}
Holding $K_{-i}$ fixed, player $i$ faces the single-player stochastic LQ problem with drift matrix $A_{-i}$, diffusion matrix $C_{-i}$, input matrices $B_i$ and $D_i$, state weight $Q_i+\sum_{j\neq i}K_j^\top R_{ij}K_j$, and input weight $R_{ii}$,
\begin{equation}
\min_{\widetilde K_i:\ (\widetilde K_i,K_{-i})\in\mathcal K}\
J_i\bigl((\widetilde K_i,K_{-i});x_0\bigr).
\label{eq:br_problem}
\end{equation}

\begin{lemma}[Best response]\label{lem:br}
Let $K\in\mathcal K$ and $i\in\mathcal I$. Then \eqref{eq:br_problem} has a unique minimizer $\widehat K_i(K_{-i})$, which does not depend on $x_0$. With $\widehat P_i(K_{-i}):=P_i\bigl((\widehat K_i(K_{-i}),K_{-i})\bigr)\in\mathbb S^n_{++}$, it satisfies
\begin{equation}
\begin{aligned}
\bigl(R_{ii}+D_i^\top\widehat P_i(K_{-i})D_i\bigr)\widehat K_i(K_{-i})
={}&B_i^\top\widehat P_i(K_{-i})\\
&+D_i^\top\widehat P_i(K_{-i})C_{-i}.
\end{aligned}
\label{eq:br_stationarity}
\end{equation}
\end{lemma}

\begin{proof}
The single-player problem is mean-square stabilizable by $K_i$, its state weight is positive definite since it is bounded from below by $Q_i\succ0$, and $R_{ii}\succ0$. Hence its stochastic algebraic Riccati equation, formed by \eqref{eq:policy-eval} for player $i$ and \eqref{eq:br_stationarity}, has a unique stabilizing solution, whose gain is optimal among all stabilizing gains \cite{aitrami_zhou_2000,Damm2004}.
\end{proof}

Lemma~\ref{lem:br} defines the best-response map $K_{-i}\mapsto\widehat K_i(K_{-i})$ of each player.
By \eqref{eq:induced_mats}, condition \eqref{eq:br_stationarity} coincides with \eqref{eq:stationarity} whenever $K_i=\widehat K_i(K_{-i})$. By Definition~\ref{def:LFNE} and the uniqueness in Lemma~\ref{lem:br}, a gain profile $K^\star\in\mathcal K$ is an equilibrium if and only if
\begin{equation}
K_i^\star=\widehat K_i(K_{-i}^\star),\qquad i\in\mathcal I.
\label{eq:br_fixed_point}
\end{equation}
IBR iterates the best-response maps of all players from the common profile $K^k$,
\begin{equation}
K^{k+1}=\mathcal T_{\mathrm{IBR}}(K^k):=\bigl(\widehat K_1(K_{-1}^k),\dots,\widehat K_N(K_{-N}^k)\bigr),
\label{eq:ibr}
\end{equation}
and, by \eqref{eq:br_fixed_point}, its fixed points are precisely the equilibria.
Each IBR step solves $N$ stochastic algebraic Riccati equations, each in turn by an inner policy iteration \cite{kleinman_1969}.
In this step, every player responds to the gains $K_{-i}^k$ of the others as if they remained fixed.

\subsection{Coupled Policy Improvement}\label{sec:coupled_improvement}

Rather than computing the exact best response $\widehat K_i(K_{-i}^k)$, CI-PI performs a single improvement step.
In \eqref{eq:br_stationarity}, CI-PI replaces the value matrix $\widehat P_i(K_{-i}^k)$ by the current policy-evaluation output $P_i(K^k)$.
It also replaces the gains of the other players in $C_{-i}$ by their updated gains $K_{-i}^{k+1}$ instead of holding them at $K_{-i}^k$.
The resulting stationarity conditions are enforced jointly across all players. Given $P^k:=P(K^k)$, the updated gains $K^{k+1}$ of all players are thus determined by the block linear system
{\small
\begin{equation}\label{eq:block_system}
\setlength{\arraycolsep}{2pt}
\underbrace{\left[\!\begin{array}{ccc}
H_{11}(P^k) & \cdots & H_{1N}(P^k)\\
\vdots & \ddots & \vdots\\
H_{N1}(P^k) & \cdots & H_{NN}(P^k)
\end{array}\!\right]}_{=:H(P^k)}
\underbrace{\left[\!\begin{array}{c}
K_1^{k+1}\\ \vdots\\ K_N^{k+1}
\end{array}\!\right]}_{=:K^{k+1}}
=
\underbrace{\left[\!\begin{array}{c}
b_1(P^k)\\ \vdots\\ b_N(P^k)
\end{array}\!\right]}_{=:b(P^k)},
\end{equation}
}%
with the blocks $H_{ii}(P)=R_{ii}+D_i^\top P_iD_i$ and
$H_{ij}(P)=D_i^\top P_iD_j$ for $i\neq j$, and
$b_i(P)=B_i^\top P_i+D_i^\top P_iC$ for any profile $P=(P_1,\dots,P_N)$ of value matrices.
We call $H(P)$ the improvement matrix.
The off-diagonal blocks of $H(P^k)$ couple the stationarity conditions of the players through their common control-dependent noise.
Block row $i$ of \eqref{eq:block_system} depends only on the value $P_i^k$ of player $i$, so that the players only need to exchange their rows.%

CI-PI takes the coupled step
\begin{equation}
K^{k+1}=\mathcal T_{\mathrm{CI}}(K^k):=H(P(K^k))^{-1}b(P(K^k)).
\label{eq:T_ci}
\end{equation}
Away from equilibria, $H(P^k)$ may be singular, since its block rows carry the values of different players.
At iterates where $H(P^k)$ is singular, $K^{k+1}=\mathcal T_{\mathrm{CI}}(K^k)$ is instead given by the block rows of \eqref{eq:block_system} with the gains of the other players held at $K_{-i}^k$, that is, by $H_{ii}(P^k)K_i^{k+1}=b_i(P^k)-\sum_{j\neq i}H_{ij}(P^k)K_j^k$ for $i\in\mathcal I$.
Since $H_{ii}(P^k)\succ0$, this step is always uniquely defined, so that $\mathcal T_{\mathrm{CI}}$ is defined on all of $\mathcal K$.
At an equilibrium $K^\star$, in contrast, nonsingularity of $H(P^\star)$ is a property of the equilibrium itself, namely that the stationarity conditions \eqref{eq:block_system} at its values $P^\star$ determine $K^\star$ uniquely.
Note that such equilibria are the usual setting for related games \cite{basar_olsder_1998,nortmann_2024}.
Accordingly, the remainder of the paper concerns these equilibria, near which CI-PI takes the coupled step by continuity.
We show in Section~\ref{sec:rates} that under growing noise, equilibria indeed have this property, that is, $H(P^\star)$ is nonsingular.

\begin{algorithm}[t]
\caption{Policy Iteration with Coupled Improvement (CI-PI)}
\label{alg:cipi}
\begin{algorithmic}[1]
\REQUIRE Stabilizing initial profile $K^0\in\mathcal K$, tolerance $\varepsilon>0$
\STATE $k\gets0$
\REPEAT
    \STATE \textbf{Policy evaluation:} $P^k\gets P(K^k)$ via \eqref{eq:policy-eval}
    \STATE \textbf{Policy improvement:} $K^{k+1}\gets\mathcal T_{\mathrm{CI}}(K^k)$ via \eqref{eq:T_ci}%
    \STATE $k\gets k+1$
\UNTIL{$\sum_{i\in\mathcal I}\|K_i^k-K_i^{k-1}\|_F<\varepsilon$}
\RETURN $K^k$
\end{algorithmic}
\end{algorithm}

\begin{proposition}[Fixed points of CI-PI]\label{prop:ci_fixed_points}
A gain profile $K\in\mathcal K$ satisfies $\mathcal T_{\mathrm{CI}}(K)=K$ if and only if $K$ is an equilibrium.
\end{proposition}

\begin{proof}
By \eqref{eq:T_ci} and the step at singular $H(P(K))$, $\mathcal T_{\mathrm{CI}}(K)=K$ holds if and only if $H(P(K))K=b(P(K))$, whether or not $H(P(K))$ is singular.
The $i$-th block row of this system is \eqref{eq:stationarity} for player $i$, and the claim follows from Proposition~\ref{prop:equilibrium}.
\end{proof}

Each iteration of CI-PI, summarized in Algorithm~\ref{alg:cipi}, requires $N$ independent Lyapunov solves and one linear solve, whereas each IBR iteration solves $N$ stochastic algebraic Riccati equations.
A stabilizing initial profile $K^0\in\mathcal K$ can be constructed from the game data by the homotopy initialization of \cite[Sec.~V]{seqPI}.
This initialization preserves mean-square stability along a continuation path and terminates after finitely many steps under Assumption~\ref{ass:MSS}.

\section{Local Convergence of CI-PI and IBR}\label{sec:local}

This section computes the local derivatives of CI-PI and IBR at an equilibrium.
The coupled equilibrium conditions may admit several equilibria, already in the deterministic special case \cite{engwerda_2005}.
The convergence of both schemes is therefore a local property that depends on the equilibrium.
Section~\ref{sec:criterion} states the local convergence criterion, and Section~\ref{sec:residuals} linearizes the residual maps on which both schemes are built.
Sections~\ref{sec:local_ibr} and~\ref{sec:local_ci} then give the derivatives of IBR and CI-PI in Propositions~\ref{prop:ibr_local} and~\ref{prop:ci_local}.

\subsection{Local Convergence Criterion}\label{sec:criterion}

CI-PI and IBR are fixed-point iterations $K^{k+1}=\mathcal T(K^k)$.
By Proposition~\ref{prop:ci_fixed_points} and \eqref{eq:br_fixed_point}, their fixed points are the equilibria.
Near such a fixed point $K^\star$, the deviation $\Delta K^k:=K^k-K^\star$ evolves to first order as $\Delta K^{k+1}\approx\D\mathcal T(K^\star)[\Delta K^k]$, where the Fr\'echet derivative $\D\mathcal T(K^\star)$ is a linear map on gain perturbations \cite{seqPI}.
Thus, its spectral radius decides whether, and how fast, the iteration converges near $K^\star$.
If $\mathcal T$ is continuously Fr\'echet differentiable near $K^\star$ and $\rho(\D\mathcal T(K^\star))<1$, the iterates converge linearly to $K^\star$ from all initial gains in a neighborhood of $K^\star$ \cite[Thms.~10.1.3, 10.1.4]{ortega_rheinboldt_1970}.
If $\D\mathcal T(K^\star)=0$, the convergence is superlinear \cite[Rem.~1]{seqPI}.
If $\rho(\D\mathcal T(K^\star))>1$, in contrast, $K^\star$ is an unstable fixed point of the iteration \cite{HirschPughShub1977}.

Throughout Sections~\ref{sec:local} and~\ref{sec:noise}, $K^\star$ denotes a fixed equilibrium, and gain perturbations $\Delta K\in\mathbb R^{m\times n}$ are stacked like gain profiles, with blocks $\Delta K_i$.

\subsection{Evaluation and Improvement Residuals}\label{sec:residuals}

Following \cite{seqPI}, we describe both schemes by two residual maps, one for policy evaluation and one for policy improvement.
Their linearizations at the equilibrium are the building blocks of the derivatives in Sections~\ref{sec:local_ibr} and~\ref{sec:local_ci}.
For $P\in\mathbb S^n$, a gain $\widetilde K_i\in\mathbb R^{m_i\times n}$ of player $i$ and a gain profile $K$, let
{\small\setlength{\mathindent}{0pt}
\begin{align}
\mathcal E_i(P,K)&:=\Lyap_K(P)+S_i(K),\label{eq:E_i}\\
\mathcal G_i(P,\widetilde K_i,K_{-i})&:=\bigl(R_{ii}+D_i^\top PD_i\bigr)\widetilde K_i-B_i^\top P-D_i^\top PC_{-i},\label{eq:G_i}
\end{align}
}%
with $C_{-i}$ from \eqref{eq:induced_mats}.
The policy evaluation residual $\mathcal E_i$ is the left-hand side of \eqref{eq:policy-eval}, so that the value $P_i(K)$ of a profile $K$ is the zero of $\mathcal E_i(\cdot,K)$, that is, $\mathcal E_i(P_i(K),K)=0$.
The improvement residual $\mathcal G_i$ is the stationarity condition of player $i$ for a given value $P$, cf. \cite[(31), (36)]{seqPI}.
By Proposition~\ref{prop:equilibrium}, $K^\star$ is an equilibrium if and only if $\mathcal G_i(P_i^\star,K_i^\star,K_{-i}^\star)=0$ for all $i\in\mathcal I$.

Both schemes compute the new gains as zeros of the improvement residuals and differ only in the arguments.
By \eqref{eq:br_stationarity}, the gains $K^{k+1}=\mathcal T_{\mathrm{IBR}}(K^k)$ of IBR, that is, the best responses $K_i^{k+1}=\widehat K_i(K_{-i}^k)$, satisfy
\begin{equation}\label{eq:G_ibr}
\mathcal G_i\bigl(P_i((K_i^{k+1},K_{-i}^k)),K_i^{k+1},K_{-i}^k\bigr)=0,\qquad i\in\mathcal I,
\end{equation}
whereas, by \eqref{eq:block_system}, the gains $K^{k+1}=\mathcal T_{\mathrm{CI}}(K^k)$ of CI-PI satisfy
\begin{equation}\label{eq:G_ci}
\mathcal G_i\bigl(P_i(K^k),K_i^{k+1},K_{-i}^{k+1}\bigr)=0,\qquad i\in\mathcal I.
\end{equation}

Differentiating \eqref{eq:E_i} and \eqref{eq:G_i} at the equilibrium gives the partial Fr\'echet derivatives, marked by $\star$, cf. \cite[App.~I]{seqPI}.
With $\Acl^\star:=\Acl(K^\star)$ and $\Ccl^\star:=\Ccl(K^\star)$, they read, for $i,j\in\mathcal I$, $Z\in\mathbb S^n$ and $\Delta K_j\in\mathbb R^{m_j\times n}$,
\begin{align}
(\D_P\mathcal E_i)^\star[Z]&=\Acl^{\star\top}Z+Z\Acl^\star+\Ccl^{\star\top}Z\Ccl^\star,\nonumber\\
(\D_{K_j}\mathcal E_i)^\star[\Delta K_j]&=-\Delta K_j^\top M_{ij}-M_{ij}^\top\Delta K_j,\label{eq:dE}\\[2pt]
(\D_P\mathcal G_i)^\star[Z]&=-B_i^\top Z-D_i^\top Z\Ccl^\star,\nonumber\\
(\D_{\widetilde K_i}\mathcal G_i)^\star[\Delta K_i]&=\bigl(R_{ii}+D_i^\top P_i^\star D_i\bigr)\Delta K_i,\label{eq:dG}\\
(\D_{K_j}\mathcal G_i)^\star[\Delta K_j]&=D_i^\top P_i^\star D_j\,\Delta K_j,\quad j\neq i,\nonumber
\end{align}
with $M_{ij}:=B_j^\top P_i^\star+D_j^\top P_i^\star\Ccl^\star-R_{ij}K_j^\star$.
The last two lines are the left multiplications by the blocks $H_{ii}(P^\star)$ and $H_{ij}(P^\star)$ of the improvement matrix in \eqref{eq:block_system}.
By \eqref{eq:stationarity}, $M_{ii}=0$, that is, the value of every player is stationary in its own gain, since this gain is optimal for its own cost.
Since $K^\star\in\mathcal K$, $(\D_P\mathcal E_i)^\star=\Lyap_{K^\star}$ is invertible \cite[Thm.~1.5.3]{Damm2004}.
By the implicit function theorem, $K\mapsto P_i(K)$ is continuously Fr\'echet differentiable near $K^\star$ \cite[Lemma~2]{seqPI}.
Its derivative $Z_i:=\D P_i(K^\star)[\Delta K]$, the change of the value of player $i$ caused by the gain perturbations of the other players, solves the Lyapunov equation
\begin{equation}\label{eq:Zi}
\Acl^{\star\top}Z_i+Z_i\Acl^\star+\Ccl^{\star\top}Z_i\Ccl^\star=\sum_{j\neq i}\bigl(\Delta K_j^\top M_{ij}+M_{ij}^\top\Delta K_j\bigr).
\end{equation}

The derivatives of both schemes are built from two linear maps on gain perturbations, the coupling $\mathcal J$ and the sensitivity $\mathcal A$ at $K^\star$, defined blockwise by
\begin{equation}\label{eq:J_A}
\begin{aligned}
(\mathcal J[\Delta K])_i&:=-H_{ii}(P^\star)^{-1}\sum_{j\neq i}H_{ij}(P^\star)\Delta K_j,\\
(\mathcal A[\Delta K])_i&:=H_{ii}(P^\star)^{-1}\bigl(B_i^\top Z_i+D_i^\top Z_i\Ccl^\star\bigr).
\end{aligned}
\end{equation}
The coupling $\mathcal J$ is the direct effect of the gains of the other players on the improvement residual of player $i$, given by $(\D_{K_j}\mathcal G_i)^\star$ in \eqref{eq:dG} and normalized by $(\D_{\widetilde K_i}\mathcal G_i)^\star=H_{ii}(P^\star)$.
The sensitivity $\mathcal A$ is their indirect effect along the chain from $(\D_{K_j}\mathcal E_i)^\star$ in \eqref{eq:dE} through the value change $Z_i$ of \eqref{eq:Zi} to $(\D_P\mathcal G_i)^\star$ in \eqref{eq:dG}, with the same normalization.
Since $M_{ii}=0$, the own gain of player $i$ has no indirect effect.
The coupling stems solely from the noise that the other players inject through the common source $W$, since it consists of the blocks $D_i^\top P_i^\star D_j$, $j\neq i$.

\subsection[Local Dynamics of IBR]{Local Dynamics of IBR}\label{sec:local_ibr}

A best response is defined only implicitly as a zero of \eqref{eq:G_ibr}.
Hence the derivative of IBR follows from the implicit function theorem.

\begin{proposition}[Derivative of IBR]\label{prop:ibr_local}
The map $\mathcal T_{\mathrm{IBR}}$ is continuously Fr\'echet differentiable near $K^\star$, and
\begin{equation}\label{eq:DT_ibr}
\D\mathcal T_{\mathrm{IBR}}(K^\star)=\mathcal J+\mathcal A.
\end{equation}
\end{proposition}

\begin{proof}
Fix $i\in\mathcal I$.
Since $M_{ii}=0$, $(\D_{K_i}\mathcal E_i)^\star=0$ by \eqref{eq:dE}.
Hence the partial derivative of the left-hand side of \eqref{eq:G_ibr} with respect to $K_i^{k+1}$ at $K^\star$ is $(\D_{\widetilde K_i}\mathcal G_i)^\star$.
This derivative is invertible by \eqref{eq:dG}, since $H_{ii}(P^\star)\succ0$.
The implicit function theorem \cite[Thm.~5.2.4]{ortega_rheinboldt_1970} yields a continuously differentiable map $K_{-i}\mapsto\widetilde K_i(K_{-i})$ near $K_{-i}^\star$ with $\widetilde K_i(K_{-i}^\star)=K_i^\star$ that solves \eqref{eq:G_ibr}.
Since $\mathcal K$ is open, $(\widetilde K_i(K_{-i}),K_{-i})\in\mathcal K$.
Hence $P_i((\widetilde K_i(K_{-i}),K_{-i}))$ is a stabilizing solution of the stochastic algebraic Riccati equation of the single-player problem \eqref{eq:br_problem}.
Unlike the coupled equations of the game, this equation has at most one stabilizing solution \cite{aitrami_zhou_2000,Damm2004}, so that $\widetilde K_i=\widehat K_i$ by Lemma~\ref{lem:br}.
Differentiating \eqref{eq:G_ibr} at $K^\star$ in the direction $\Delta K$ and using \eqref{eq:dE}, \eqref{eq:dG} and \eqref{eq:Zi} gives, with $\delta K:=\D\mathcal T_{\mathrm{IBR}}(K^\star)[\Delta K]$,
\[
H_{ii}(P^\star)\,\delta K_i=B_i^\top Z_i+D_i^\top Z_i\Ccl^\star-\sum_{j\neq i}H_{ij}(P^\star)\Delta K_j .
\]
By \eqref{eq:J_A}, $\delta K_i$ is the $i$-th block of $(\mathcal J+\mathcal A)[\Delta K]$.
\end{proof}

The derivative of IBR involves only the inverses of the diagonal blocks $H_{ii}(P^\star)$ of the improvement matrix.
Each player corrects its own stationarity residual but takes the noise injected by the other players as given.
This noise therefore enters the derivative of IBR additively, as the coupling $\mathcal J$.

\begin{remark}[Exact best responses]\label{rem:decoupled}
The derivative \eqref{eq:DT_ibr} of IBR coincides with the derivative of the decoupled policy iteration of \cite[Prop.~3]{seqPI}, in which all players simultaneously perform a single improvement step with the gains of the others held fixed.
The two derivatives agree because $M_{ii}=0$, that is, the value of each player is, to first order, insensitive to its own gain.
Since the fixed points of both iterations are the equilibria, IBR and the decoupled policy iteration have the same local rate at every equilibrium.
The inner iterations of IBR, which compute exact best responses, thus increase the effort per iteration without improving the local rate.
This conclusion does not rely on control-dependent noise and holds verbatim in the deterministic special case.
\end{remark}

\subsection[Local Dynamics of CI-PI]{Local Dynamics of CI-PI}\label{sec:local_ci}

Unlike a best response, the new gains of CI-PI solve the linear system \eqref{eq:block_system}, so that the derivative of CI-PI follows by differentiating \eqref{eq:G_ci} directly.

\begin{proposition}[Derivative of CI-PI]\label{prop:ci_local}
The map $\mathcal T_{\mathrm{CI}}$ is continuously Fr\'echet differentiable near $K^\star$, and
\begin{equation}\label{eq:DT_ci}
\D\mathcal T_{\mathrm{CI}}(K^\star)=(I-\mathcal J)^{-1}\mathcal A.
\end{equation}
\end{proposition}

\begin{proof}
Since $K\mapsto P(K)$ is continuously Fr\'echet differentiable near $K^\star$ and $H(P^\star)$ is nonsingular, $H(P(K))$ is nonsingular near $K^\star$, and $\mathcal T_{\mathrm{CI}}(K)=H(P(K))^{-1}b(P(K))$ is continuously Fr\'echet differentiable there.
Differentiating \eqref{eq:G_ci} at $K^\star$ in the direction $\Delta K$ and using \eqref{eq:dG} and \eqref{eq:Zi} gives, with $\delta K:=\D\mathcal T_{\mathrm{CI}}(K^\star)[\Delta K]$,
\[
H_{ii}(P^\star)\,\delta K_i+\sum_{j\neq i}H_{ij}(P^\star)\,\delta K_j=B_i^\top Z_i+D_i^\top Z_i\Ccl^\star
\]
for all $i\in\mathcal I$, that is, $(I-\mathcal J)\,\delta K=\mathcal A[\Delta K]$.
Since $H(P^\star)$ is nonsingular, $I-\mathcal J$ is invertible, which gives \eqref{eq:DT_ci}.
\end{proof}

The coupling $\mathcal J$ therefore enters its derivative only through the factor $(I-\mathcal J)^{-1}$ of the sensitivity $\mathcal A$, and $\D\mathcal T_{\mathrm{CI}}(K^\star)$ vanishes whenever $\mathcal A$ vanishes, however strong the coupling is.
By \eqref{eq:J_A}, $H(P^\star)$ is nonsingular if and only if $1$ is not an eigenvalue of $\mathcal J$.%

The local dynamics of both schemes are governed by the same two maps $\mathcal J$ and $\mathcal A$ and differ only in how the coupling enters.
By Section~\ref{sec:criterion}, IBR thus converges locally if $\rho(\mathcal J+\mathcal A)<1$, and CI-PI if $\rho((I-\mathcal J)^{-1}\mathcal A)<1$.
Propositions~\ref{prop:ibr_local} and~\ref{prop:ci_local} reduce the second requirement of Problem~\ref{prob:cipi} to the behavior of $\mathcal J$ and $\mathcal A$, which Section~\ref{sec:noise} analyzes under growing noise.

\section{Local Rates under Growing Noise}\label{sec:noise}
\providecommand{\DXD}{D^\top XD}

This section analyzes the local rates of CI-PI and IBR at equilibria under growing noise.
To study the effect of different noise levels, we replace $C$ and $D_i$ by $sC$ and $sD_i$ with a noise level $s\ge0$ and write $\mathcal K(s)$ to indicate the dependence on $s$.
This scaling does not restrict the class of games, since $s=1$ recovers the original game.
As the noise level $s$ grows, the coupling blocks $s^2D_i^\top P_i^\star D_j$ of the improvement matrix grow, and the set $\mathcal K(s)$ shrinks \cite[Cor.~3.2.4]{Damm2004}.
An equilibrium may thus come close to losing mean-square stability.
We show that both effects are governed by the slowest decaying mode of the second moment $\mathbb E[x(t)x(t)^\top]$ of the closed loop and its decay rate, the stability margin.
Section~\ref{sec:margin} introduces this margin, and Section~\ref{sec:rates} derives the resulting local rates of CI-PI and IBR in Theorem~\ref{thm:rates} and Corollary~\ref{cor:geometry}.

\subsection{Stability Margin and Critical Mode}\label{sec:margin}

To describe equilibria close to the boundary of $\mathcal K$, we first quantify the distance to this boundary by the decay rate of the slowest mode of the second moment and then split the values of the players along this mode.

For $K\in\mathcal K$, the operator $\Lyap_K$ is resolvent positive with respect to $\mathbb S^n_+$ \cite[Ex.~3.2.2, Cor.~3.2.4]{Damm2004}.
Its spectral abscissa, the largest real part of its spectrum, is therefore itself an eigenvalue, with eigenvectors in $\mathbb S^n_+\setminus\{0\}$ for $\Lyap_K$ and for its adjoint $\Lyap_K^*(Y)=\Acl(K)Y+Y\Acl(K)^\top+\Ccl(K)Y\Ccl(K)^\top$ \cite[Thm.~3.2.3]{Damm2004}.
We call
\begin{equation}\label{eq:margin}
\kappa(K):=-\max\RRe\spec(\Lyap_K)
\end{equation}
the MSS margin of $K$.
It is the decay rate of the slowest mode of the second moment $\mathbb E[x(t)x(t)^\top]$, which evolves according to $\Lyap_K^*$ \cite[Thm.~1.4.3]{Damm2004}.
Moreover, $K\in\mathcal K$ if and only if $\kappa(K)>0$ \cite[Thm.~1.5.3]{Damm2004}.
The margin thus quantifies the stability reserve of the closed loop.

Let $\Lyap:=\Lyap_{K^\star}$ and $\kappa:=\kappa(K^\star)$, and let $-\kappa$ be an algebraically simple eigenvalue of $\Lyap$ with eigenvectors $X,Y\in\mathbb S^n_+$ of $\Lyap$ and $\Lyap^*$.
The range of $\Lyap+\kappa I$ is the orthogonal complement of the kernel $\operatorname{span}\{Y\}$ of its adjoint $\Lyap^*+\kappa I$, that is, the hyperplane $Y^\perp:=\{Z\in\mathbb S^n:\langle Y,Z\rangle=0\}$.
Since the eigenvalue is simple, $X$ does not lie in this range, as otherwise $(\Lyap+\kappa I)Z=X$ would have a solution $Z$ and $-\kappa$ would not be simple.
Hence $\langle Y,X\rangle>0$, and we normalize $\|X\|_F=1=\langle Y,X\rangle$ and call $X$ the critical mode.
The space $\mathbb S^n$ is the direct sum of the line spanned by $X$, on which $\Lyap$ acts as $-\kappa$, and of the invariant hyperplane $Y^\perp$.
Consequently, for all $Z\in\mathbb S^n$,
\begin{equation}\label{eq:split}
\Lyap^{-1}(Z)=-\frac{\langle Y,Z\rangle}{\kappa}\,X+\R(Z),
\end{equation}
where $\R(Z):=\Lyap^{-1}(Z-\langle Y,Z\rangle X)$ is the inverse of $\Lyap$ on $Y^\perp$, applied to the projection of $Z$ onto $Y^\perp$ along $X$.
Unlike the first term, $\R$ does not involve $1/\kappa$.

Applied to the value equations, \eqref{eq:split} isolates the part of every value that grows as the margin shrinks.
This part enters the coupling blocks $D_i^\top P_i^\star D_j$ of the improvement matrix and determines the local rates in Section~\ref{sec:rates}.

\begin{lemma}[Values at small margin]\label{lem:critical}
For $i\in\mathcal I$, let $c_i:=\langle Y,S_i(K^\star)\rangle$. Then $c_i\ge\lambda_{\min}(Q_i)$ and
\begin{equation}\label{eq:value_split}
P_i^\star=\frac{c_i}{\kappa}\,X-\R\bigl(S_i(K^\star)\bigr).
\end{equation}
\end{lemma}

\begin{proof}
By \eqref{eq:policy-eval}, $P_i^\star=-\Lyap^{-1}(S_i(K^\star))$, and \eqref{eq:split} gives \eqref{eq:value_split}.
Since $\|X\|_F=1=\langle Y,X\rangle$, the Cauchy--Schwarz inequality gives $\tr Y\ge\|Y\|_F\ge1$, and $S_i(K^\star)\succeq Q_i$ and $Y\succeq0$ give $c_i\ge\lambda_{\min}(Q_i)\tr Y\ge\lambda_{\min}(Q_i)$.
\end{proof}

For small $\kappa$, the first term in \eqref{eq:value_split} dominates the value of every player.
Like the splitting itself, the critical mode $X$ depends on the equilibrium.

\begin{remark}[Small margin]\label{rem:small_margin}
Equilibria with a small margin occur under control-dependent noise.
For example, the values $P_i^0$ at $K=0$ do not depend on the input matrices, and \eqref{eq:stationarity} is linear in $B_i$ there.
Hence $K^\star=0$ is an equilibrium for $B_i:=-(P_i^0)^{-1}C^\top P_i^0D_i$ whenever $0\in\mathcal K$ (Proposition~\ref{prop:equilibrium}).
Its margin depends on $A$ and $C$ alone and vanishes as $C$ is scaled up to the level at which $K=0$ ceases to be mean-square stabilizing.
\end{remark}

\subsection{Local Rates at Small Margin}\label{sec:rates}

By Lemma~\ref{lem:critical}, the part of the values that grows like $1/\kappa$ lies along the critical mode and is common to all players up to the scalar weights $c_i$.
This common structure determines the local rates at small margin.
We first derive its effect on the coupling $\mathcal J$ and on the sensitivity $\mathcal A$ of \eqref{eq:J_A}.
Theorem~\ref{thm:rates} then combines both in the local derivatives \eqref{eq:DT_ibr} and \eqref{eq:DT_ci} of IBR and CI-PI.
Here and below, $H:=H(P^\star)$, and $M_{\mathrm d}$ and $M_{\mathrm o}$ denote the block-diagonal and off-diagonal parts of a block matrix $M$, so that $\mathcal J[\Delta K]=-H_{\mathrm d}^{-1}H_{\mathrm o}\Delta K$.
Matrices $M\in\mathbb R^{m\times m}$ act on $\Delta K$ by left multiplication, with the same eigenvalues and norm as $M$.
Expressions such as $\D\mathcal T_{\mathrm{IBR}}(K^\star)-M$ are understood in this sense.

We begin with the coupling $\mathcal J$.
Inserting \eqref{eq:value_split} into the blocks of $H$ in \eqref{eq:block_system} gives
\begin{equation}\label{eq:H_scaled}
\frac{\kappa}{c_i}\,H_{ij}=D_i^\top XD_j+\kappa F_{ij}
\end{equation}
with $F_{ii}:=\bigl(R_{ii}-D_i^\top\R(S_i(K^\star))D_i\bigr)/c_i$ and $F_{ij}:=-D_i^\top\R(S_i(K^\star))D_j/c_i$ for $j\neq i$.
In the $i$-th block row of $\mathcal J$, the scalar factor $\kappa/c_i$ cancels, since $H_{ii}^{-1}H_{ij}=(\frac{\kappa}{c_i}H_{ii})^{-1}\frac{\kappa}{c_i}H_{ij}$.
Hence, by \eqref{eq:J_A} and \eqref{eq:H_scaled},
\[
\begin{aligned}
(\mathcal J[\Delta K])_i={}&-\bigl(D_i^\top XD_i+\kappa F_{ii}\bigr)^{-1}\\
&\cdot\sum_{j\neq i}\bigl(D_i^\top XD_j+\kappa F_{ij}\bigr)\Delta K_j .
\end{aligned}
\]
In this representation, the coupling is built from two parts.
The first part, $D_i^\top XD_j$, involves only the critical mode and the noise channels and does not carry the factor $\kappa$.
The second part, $\kappa F_{ij}$, collects the control weights and the parts of the values off the critical mode and is multiplied by the margin.
The first part alone defines the limit coupling.

\begin{definition}[Limit coupling]\label{def:limit_coupling}
Let the critical eigenvalue $-\kappa$ of $\Lyap$ be algebraically simple with critical mode $X$, and let $D_i^\top XD_i\succ0$ for all $i\in\mathcal I$.
The limit coupling at $K^\star$ is
\begin{equation}\label{eq:JX}
\mathcal J_X:=-(\DXD)_{\mathrm d}^{-1}(\DXD)_{\mathrm o}\in\mathbb R^{m\times m}.
\end{equation}
\end{definition}

The limit coupling is the coupling $\mathcal J$ of \eqref{eq:J_A} with all values $P_i^\star$ replaced by the critical mode $X$ and the control weights $R_{ii}$ omitted.
The second part, in contrast, vanishes with the margin.
If the margin shrinks, for instance under growing noise, and $D_i^\top XD_i\succeq\omega I$, the coupling $\mathcal J$ therefore approaches the limit coupling.

We now turn to the sensitivity $\mathcal A$.
Its terms $B_i^\top Z_i+D_i^\top Z_i\Ccl^\star$ involve the value changes $Z_i$ of \eqref{eq:Zi}.
By \eqref{eq:Zi} and \eqref{eq:split}, these changes have a component along the critical mode $X$ that grows like $1/\kappa$.
The next lemma shows that their effect on the improvement steps nevertheless vanishes with the margin.

\begin{lemma}[Vanishing sensitivity]\label{lem:sensitivity}
Let $r,\omega>0$, and consider equilibria $K^\star$ whose critical eigenvalue $-\kappa$ is algebraically simple, with $X$, $Y$ and $\R$ as in Section~\ref{sec:margin}, such that $\|K^\star\|\le r$, $\|Y\|_F\le r$ and $\|\R\|\le r$, which keep the critical eigenvalue uniformly simple, and $D_i^\top XD_i\succeq\omega I$ for all $i\in\mathcal I$.
There are a threshold $\bar\kappa>0$ and a constant $\sigma>0$, depending only on $r$, $\omega$ and on bounds for the system and cost matrices, such that every such equilibrium with $\kappa\le\bar\kappa$ satisfies%
\begin{equation}\label{eq:A_bound}
\|\mathcal A\|\le\sigma\kappa
\end{equation}
for the sensitivity $\mathcal A$ of \eqref{eq:J_A}.
\end{lemma}

\begin{proof}
All constants below depend only on $r$, $\omega$ and the bounds for the system and cost matrices.
By \eqref{eq:J_A}, $(\mathcal A[\Delta K])_i=H_{ii}^{-1}\bigl(B_i^\top Z_i+D_i^\top Z_i\Ccl^\star\bigr)$.
We show that the term $B_i^\top Z_i+D_i^\top Z_i\Ccl^\star$ is bounded by a constant times $\|\Delta K\|_F$, although $Z_i$ may grow like $1/\kappa$, and that $H_{ii}^{-1}$ is bounded by a constant times $\kappa$.

(Auxiliary bounds)
The weights $c_i=\langle Y,S_i(K^\star)\rangle$ of Lemma~\ref{lem:critical} satisfy $\lambda_{\min}(Q_i)\le c_i\le\|Y\|_F\|S_i(K^\star)\|_F$ by Lemma~\ref{lem:critical} and the Cauchy--Schwarz inequality, where the upper bound is bounded since $\|Y\|_F\le r$ and $S_i(K^\star)=Q_i+\sum_{j\in\mathcal I}K_j^{\star\top}R_{ij}K_j^\star$ with $\|K^\star\|\le r$.
Since $M_{ii}=0$, the stationarity condition \eqref{eq:stationarity} reads $R_{ii}K_i^\star=B_i^\top P_i^\star+D_i^\top P_i^\star\Ccl^\star$.
Inserting \eqref{eq:value_split} splits its right-hand side into a part along $X$, which carries the factor $c_i/\kappa$, and a remainder involving $\R(S_i(K^\star))$.
Collecting the remainder with $R_{ii}K_i^\star$ and multiplying by $\kappa/c_i$ gives%
\begin{equation}\label{eq:incentive}
B_i^\top X+D_i^\top X\Ccl^\star=\frac{\kappa}{c_i}\,w_i
\end{equation}
with $w_i:=R_{ii}K_i^\star+B_i^\top\R(S_i(K^\star))+D_i^\top\R(S_i(K^\star))\Ccl^\star$.
Inserting \eqref{eq:value_split} into the definition of $M_{ij}$ in Section~\ref{sec:residuals} and using \eqref{eq:incentive} for player $j$ expresses $M_{ij}$ without the factor $1/\kappa$,%
\[
\begin{aligned}
M_{ij}={}&\frac{c_i}{c_j}\,w_j-R_{ij}K_j^\star\\
&-B_j^\top\R(S_i(K^\star))-D_j^\top\R(S_i(K^\star))\Ccl^\star .
\end{aligned}
\]
Since $c_i$ is bounded from above and below, and $\|K^\star\|\le r$ and $\|\R\|\le r$, both $w_i$ and $M_{ij}$ are bounded.

(Bounds on $B_i^\top Z_i+D_i^\top Z_i\Ccl^\star$)
By \eqref{eq:Zi}, $Z_i=\Lyap^{-1}(V_i)$ with the right-hand side $V_i:=\sum_{j\neq i}(\Delta K_j^\top M_{ij}+M_{ij}^\top\Delta K_j)$.
Since $M_{ij}$ is bounded, $V_i$ is bounded by a constant times $\|\Delta K\|_F$.
By \eqref{eq:split}, $Z_i=-\frac{\langle Y,V_i\rangle}{\kappa}X+\R(V_i)$, in which only the first term carries the factor $1/\kappa$.
Inserting this into $B_i^\top Z_i+D_i^\top Z_i\Ccl^\star$ and using \eqref{eq:incentive} for the first part gives
\[
\begin{aligned}
B_i^\top Z_i+D_i^\top Z_i\Ccl^\star={}&-\frac{\langle Y,V_i\rangle}{c_i}\,w_i\\
&+B_i^\top\R(V_i)+D_i^\top\R(V_i)\Ccl^\star,
\end{aligned}
\]
in which the factor $1/\kappa$ has cancelled.
Since $c_i$ is bounded from below, $|\langle Y,V_i\rangle|\le\|Y\|_F\|V_i\|_F$, and $w_i$ and $\R$ are bounded, this term is bounded by a constant times $\|\Delta K\|_F$.%

(Bounds on $H_{ii}^{-1}$)
Let $\Gamma:=\blkdiag(c_1I_{m_1},\dots,c_NI_{m_N})$, and let $F$ be the block matrix with the blocks $F_{ij}$ of \eqref{eq:H_scaled}, so that \eqref{eq:H_scaled} reads $\kappa\Gamma^{-1}H=\DXD+\kappa F$.
By the auxiliary bounds, $\Gamma^{-1}$ and $F$ are bounded, say $\|F\|\le f$.
We set $\bar\kappa:=\omega/(2f)$, which depends only on $r$, $\omega$ and the bounds for the system and cost matrices, and let $\kappa\le\bar\kappa$.
Then $\|\kappa F_{\mathrm d}\|\le\omega/2$, so that $(\DXD)_{\mathrm d}\succeq\omega I$ gives $\|((\DXD)_{\mathrm d}+\kappa F_{\mathrm d})^{-1}\|\le2/\omega$.
Since $\Gamma$ is block diagonal, $H_{\mathrm d}^{-1}=\kappa((\DXD)_{\mathrm d}+\kappa F_{\mathrm d})^{-1}\Gamma^{-1}$ is bounded by a constant times $\kappa$.

Together with \eqref{eq:J_A}, both bounds give $\|\mathcal A\|\le\sigma\kappa$.
\end{proof}

Lemma~\ref{lem:sensitivity} rests on the equilibrium property.
The stationarity condition \eqref{eq:incentive} forces the incentive $B_i^\top X+D_i^\top X\Ccl^\star$ of every player along the critical mode to be of order $\kappa$, and this factor cancels the factor $1/\kappa$ in $Z_i$.
The following theorem combines both results in \eqref{eq:DT_ibr} and \eqref{eq:DT_ci}.
Its constants depend only on $r$, $\omega$ and the bounds for the system and cost matrices, and hence not on the particular equilibrium.

\begin{theorem}[Local rates at small margin]\label{thm:rates}
Let $r$, $\omega$ and $\bar\kappa$ be as in Lemma~\ref{lem:sensitivity}.
There is $c>0$, depending only on $r$, $\omega$ and on bounds for the system and cost matrices, such that every equilibrium $K^\star$ considered there with $\kappa\le\bar\kappa$ and $\DXD\succeq\omega I$ satisfies the following two statements.
\begin{enumerate}
\item[(i)] $\|\D\mathcal T_{\mathrm{IBR}}(K^\star)-\mathcal J_X\|\le c\,\kappa$, that is, the local behavior of IBR is governed by the limit coupling $\mathcal J_X$ of the noise channels according to Definition~\ref{def:limit_coupling};
\item[(ii)] $\|\D\mathcal T_{\mathrm{CI}}(K^\star)\|\le c\,\kappa$, that is, CI-PI converges locally with a linear rate of at most $c\,\kappa$ whenever $c\,\kappa<1$ and approaches superlinear convergence as the margin vanishes.
\end{enumerate}
\end{theorem}

\begin{proof}
All constants below depend only on $r$, $\omega$ and the bounds for the system and cost matrices.

(Scaling)
By \eqref{eq:H_scaled}, scaling the $i$-th block row of $H$ by $\kappa/c_i$ gives $\kappa\Gamma^{-1}H=\DXD+\kappa F$, with $\Gamma$ and $F$ as in the proof of Lemma~\ref{lem:sensitivity}.
Both statements rest on this scaled form.
After scaling, $H$ thus differs from $\DXD$ only by $\kappa F$, which vanishes with the margin since $F$ is bounded.
Since $\DXD\succeq\omega I$ and $\|\kappa F\|\le\omega/2$ for $\kappa\le\bar\kappa$, the inverses of $\DXD+\kappa F$ and of its block diagonal $(\DXD)_{\mathrm d}+\kappa F_{\mathrm d}$ are bounded by $2/\omega$.
Moreover, $\|\DXD\|\le\|D\|^2$ since $\|X\|_F=1$.
The block-diagonal factor $\kappa^{-1}\Gamma$ cancels in $\mathcal J=-H_{\mathrm d}^{-1}H_{\mathrm o}$, so that
\[
\mathcal J=-\bigl((\DXD)_{\mathrm d}+\kappa F_{\mathrm d}\bigr)^{-1}\bigl((\DXD)_{\mathrm o}+\kappa F_{\mathrm o}\bigr).
\]
Hence $\mathcal J$ is the limit coupling $\mathcal J_X$ with $\DXD$ replaced by $\DXD+\kappa F$.

(Statement (i))
By \eqref{eq:DT_ibr}, $\D\mathcal T_{\mathrm{IBR}}(K^\star)-\mathcal J_X=(\mathcal J-\mathcal J_X)+\mathcal A$, where $\|\mathcal A\|\le\sigma\kappa$ by Lemma~\ref{lem:sensitivity}.
Since $\mathcal J$ and $\mathcal J_X=-(\DXD)_{\mathrm d}^{-1}(\DXD)_{\mathrm o}$ differ only by the perturbation $\kappa F$, factoring out $((\DXD)_{\mathrm d}+\kappa F_{\mathrm d})^{-1}$ and using $((\DXD)_{\mathrm d}+\kappa F_{\mathrm d})(\DXD)_{\mathrm d}^{-1}(\DXD)_{\mathrm o}=(\DXD)_{\mathrm o}+\kappa F_{\mathrm d}(\DXD)_{\mathrm d}^{-1}(\DXD)_{\mathrm o}$ gives%
\[
\begin{aligned}
\mathcal J-\mathcal J_X&=\kappa\,((\DXD)_{\mathrm d}+\kappa F_{\mathrm d})^{-1}\\
&\quad\bigl(F_{\mathrm d}(\DXD)_{\mathrm d}^{-1}(\DXD)_{\mathrm o}-F_{\mathrm o}\bigr).
\end{aligned}
\]
All factors after $\kappa$ are bounded by the scaling step, since also $(\DXD)_{\mathrm d}\succeq\omega I$ and $\|F\|\le f$, and (i) follows.

\emergencystretch=1.5em
(Statement (ii))
By the scaling step, $H$ is nonsingular, so that $\mathcal T_{\mathrm{CI}}$ is continuously Fr\'echet differentiable near $K^\star$ by Proposition~\ref{prop:ci_local}.
The factor $\kappa^{-1}\Gamma$ likewise cancels in $I-\mathcal J=H_{\mathrm d}^{-1}H$, so that $(I-\mathcal J)^{-1}=(\DXD+\kappa F)^{-1}((\DXD)_{\mathrm d}+\kappa F_{\mathrm d})$ is bounded.
Hence \eqref{eq:DT_ci} and Lemma~\ref{lem:sensitivity} give $\|\D\mathcal T_{\mathrm{CI}}(K^\star)\|\le\|(I-\mathcal J)^{-1}\|\,\|\mathcal A\|\le c\,\kappa$.
\end{proof}

\begin{remark}[Acceleration of CI-PI]\label{rem:acceleration}
Theorem~\ref{thm:rates}(ii) may appear counterintuitive.
The more the noise drives the equilibrium towards losing mean-square stability, the faster CI-PI converges, although the values $P_i^\star$ grow like $1/\kappa$.
The reason is that the values align.
By \eqref{eq:value_split}, the scaled values $\frac{\kappa}{c_i}P_i^\star$ of all players differ from the same matrix $X$ by terms of order $\kappa$, irrespective of their costs, which enter $X$ only through the equilibrium.
Intuitively, the common critical mode gains weight in the cost of every player as the margin shrinks.
This holds for every player, since each state weight $Q_i$ is positive definite and Lemma~\ref{lem:critical} thus gives each player a weight $c_i\ge\lambda_{\min}(Q_i)>0$ on this mode.
Although the game remains noncooperative, the players increasingly share an interest in keeping the closed loop mean-square stable.
After the block row of player $i$ is scaled by $\kappa/c_i$, the improvement matrix thus approaches the common matrix $D^\top XD$.
CI-PI inverts this matrix exactly, while the remaining sensitivity $\mathcal A$ vanishes by Lemma~\ref{lem:sensitivity}.
IBR, in contrast, retains the coupling, which approaches the limit coupling $\mathcal J_X$ by Theorem~\ref{thm:rates}(i).
\end{remark}

Since the rate of CI-PI vanishes regardless of the coupling, the remaining comparison concerns IBR alone.
What $\mathcal J_X$ expresses, and when it lets IBR converge, is made precise next.

\begin{remark}[Geometric interpretation]\label{rem:geometry}
The limit coupling $\mathcal J_X$ depends only on the geometry of the noise channels of the players, weighted by the critical mode.
At small margin, the local rate of IBR is therefore set by how much these channels overlap.
If they are mutually orthogonal, $\mathcal J_X=0$ and the coupling disappears.
Overlapping channels, in contrast, couple the players, with a strength set by the angles between the channels.
To see this, note that $\DXD=(X^{1/2}D)^\top(X^{1/2}D)$ is the Gram matrix of the noise channels $X^{1/2}D_i$.
Its diagonal blocks $D_i^\top XD_i$ measure the strength of each channel along the critical mode, and its off-diagonal blocks $D_i^\top XD_j$ measure how much two channels overlap.
The matrix $I-\mathcal J_X=(\DXD)_{\mathrm d}^{-1}\DXD$ is similar to $(\DXD)_{\mathrm d}^{-1/2}\DXD(\DXD)_{\mathrm d}^{-1/2}$, which normalizes the channels.
The singular values of its off-diagonal blocks are the cosines of the principal angles between the noise channels, that is, between the subspaces they span, which generalize the angle between two vectors \cite{golub_vanloan_2013}.
Corollary~\ref{cor:geometry} makes the resulting rate of IBR explicit.
\end{remark}

\begin{corollary}[Rate of IBR]\label{cor:geometry}
Under the hypotheses of Theorem~\ref{thm:rates}, the spectrum of the limit coupling $\mathcal J_X$ lies in $[1-N,1)$, and there is $c'>0$, depending only on $r$, $\omega$ and the bounds for the system and cost matrices, such that
\[
\bigl|\rho(\D\mathcal T_{\mathrm{IBR}}(K^\star))-\rho(\mathcal J_X)\bigr|\le c'\kappa .
\]
In particular, two cases arise.
\begin{enumerate}
\item[(a)] For $N=2$, $\rho(\mathcal J_X)=\cos\vartheta_1<1$ with the smallest principal angle $\vartheta_1$ between the two noise channels.
Hence IBR converges locally to $K^\star$ at small margin, namely whenever $\kappa<(1-\cos\vartheta_1)/c'$.
\item[(b)] If $\rho(\mathcal J_X)>1$, which requires $N\ge3$ by (a), then IBR diverges locally at small margin.
More precisely, $K^\star$ is an unstable fixed point of IBR whenever $\kappa<(\rho(\mathcal J_X)-1)/c'$.
\end{enumerate}
\end{corollary}

\begin{proof}
\emergencystretch=1.5em
(Spectrum of $\mathcal J_X$)
Let $U_i:=X^{1/2}D_i(D_i^\top XD_i)^{-1/2}$ be the noise channels of Remark~\ref{rem:geometry}, normalized to orthonormal columns, and let $U:=[U_1\ \cdots\ U_N]$.
Then $(\DXD)_{\mathrm d}^{-1/2}\DXD(\DXD)_{\mathrm d}^{-1/2}=U^\top U$ is their Gram matrix \cite{HornJohnson2012}, and $U^\top U\succ0$ since $\DXD\succeq\omega I$.
With an eigendecomposition $U^\top U=Q\Lambda Q^\top$ and $V:=(\DXD)_{\mathrm d}^{-1/2}Q$, \eqref{eq:JX} gives $\mathcal J_X=V(I-\Lambda)V^{-1}$.
Its eigenvalues $1-\lambda$, $\lambda\in\spec(U^\top U)$, thus lie in $[1-N,1)$, since $\lambda>0$ and $\|Ux\|\le\sum_i\|U_ix_i\|=\sum_i\|x_i\|\le\sqrt N\|x\|$ for $x=(x_1,\dots,x_N)$.

(Perturbation bound)
On $\mathbb R^{m\times n}$, $\mathcal J_X$ acts by left multiplication, with the eigenvector matrix $V$ and each eigenvalue repeated $n$ times.
Let $E:=\D\mathcal T_{\mathrm{IBR}}(K^\star)-\mathcal J_X$.
By the Bauer--Fike theorem \cite[Thm.~IIIa]{bauer_fike_1960}, every eigenvalue of $\mathcal J_X+\tau E$, $\tau\in[0,1]$, lies within $\varepsilon:=\|V\|\|V^{-1}\|\|E\|$ of an eigenvalue of $\mathcal J_X$, so that $\rho(\D\mathcal T_{\mathrm{IBR}}(K^\star))\le\rho(\mathcal J_X)+\varepsilon$.
Conversely, these eigenvalues depend continuously on $\tau$ \cite[Sec.~2.1]{ortega_rheinboldt_1970}, so none of them leaves the connected component of the union of these discs in which it starts.
The component around an eigenvalue of modulus $\rho(\mathcal J_X)$ has diameter at most $2m\varepsilon$ and contains an eigenvalue of $\D\mathcal T_{\mathrm{IBR}}(K^\star)$, which gives $\rho(\D\mathcal T_{\mathrm{IBR}}(K^\star))\ge\rho(\mathcal J_X)-2m\varepsilon$.
Since $\|V\|\|V^{-1}\|=\|(\DXD)_{\mathrm d}^{-1/2}\|\|(\DXD)_{\mathrm d}^{1/2}\|\le\omega^{-1/2}\|D\|$ and $\|E\|\le c\,\kappa$ by Theorem~\ref{thm:rates}(i), both bounds give the claim.

(Cases (a) and (b))
For $N=2$, $U^\top U$ has the eigenvalues $1\pm\sigma_l$ with the singular values $\sigma_l$ of $U_1^\top U_2$ and, if $m_1\neq m_2$, the eigenvalue $1$ with multiplicity $|m_1-m_2|$.
The $\sigma_l$ are the cosines of the principal angles between the noise channels \cite{golub_vanloan_2013}, and $\sigma_l<1$ since $U^\top U\succ0$.
Hence $\rho(\mathcal J_X)=\cos\vartheta_1$.
In both (a) and (b), the bound and the criterion of Section~\ref{sec:criterion} give the consequence.
\end{proof}

At small margin, IBR thus converges for two players, at a rate set by the angle between their noise channels, but may diverge for three or more.
For instance, let all $N$ players have scalar inputs, and let all pairs of normalized channels overlap equally, $U_i^\top U_j=\gamma\in(0,1)$ for $i\neq j$.
Then $U^\top U=(1-\gamma)I+\gamma\mathbf 1\mathbf 1^\top$ with the vector $\mathbf 1\in\mathbb R^N$ of ones, so that $\mathcal J_X$ has the eigenvalues $\gamma$ and $-(N-1)\gamma$, and $\rho(\mathcal J_X)=(N-1)\gamma$.
By Corollary~\ref{cor:geometry}(b), IBR becomes locally divergent at small margin as soon as $\gamma>1/(N-1)$, for three players already at pairwise angles below $60^\circ$.

\begin{figure*}[t]
\centering
\includegraphics[width=\textwidth]{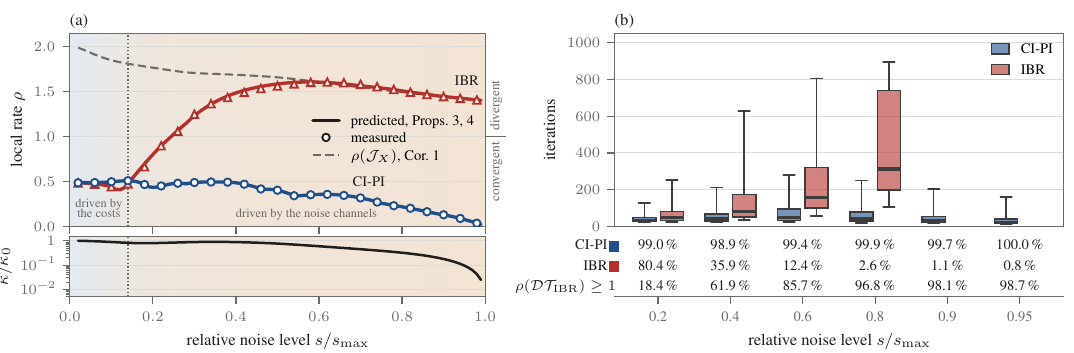}
\caption{Three-player games under parametric uncertainty ($n=10$, $m_i=2$) as the noise level $s$ approaches $s_{\max}$. (a)~One game: local rates of Propositions~\ref{prop:ibr_local} and~\ref{prop:ci_local}, measured rates, and margin $\kappa$ (bottom); background: from blue to orange as the share $\rho(\mathcal J)/(\rho(\mathcal J)+\rho(\mathcal A))$ of the coupling grows from $0$ to $1$, dotted: $\rho(\mathcal J)=\rho(\mathcal A)$, dashed: $\rho(\mathcal J_X)$. (b)~$1000$ games: iterations of the successful runs (boxes: quartiles and median, whiskers: $5$--$95\,\%$, no box for fewer than ten runs), success rates, and share of the equilibria that repel IBR.}
\label{fig:numerics}
\end{figure*}

\subsection{Regime Change}\label{sec:regime}

Together, Propositions~\ref{prop:ibr_local} and~\ref{prop:ci_local}, Theorem~\ref{thm:rates} and Corollary~\ref{cor:geometry} describe how the local dynamics change under growing noise.
At low noise, $H\approx R$ and $\mathcal J\approx0$.
Both schemes then inherit the rate of $\mathcal A$, which is set by the costs.
As the noise grows, the noise prices $D_i^\top P_i^\star D_i\approx\frac{c_i}{\kappa}D_i^\top XD_i$ outgrow the control weights $R_{ii}$.
Consequently, $\mathcal A$ vanishes with the margin (Lemma~\ref{lem:sensitivity}), and $\mathcal J$ approaches the limit coupling $\mathcal J_X$.
The local dynamics thus change from a regime driven by the costs to a regime driven by the geometry of the noise channels.
In this regime, CI-PI approaches superlinear convergence, whereas the rate of IBR approaches $\rho(\mathcal J_X)$.
With Propositions~\ref{prop:ci_fixed_points}, \ref{prop:ibr_local} and~\ref{prop:ci_local}, these results solve Problem~\ref{prob:cipi}.

\section{Numerical Examples}\label{sec:numerics}

This section shows that the regime change described in Section~\ref{sec:regime} already sets in at moderate noise levels and that it decides the behavior of the actual iterations.
As an example of games with multiplicative noise, we consider games under parametric uncertainty, in which $C$ and $D_i$ model random perturbations of $A$ and $B_i$ \cite[Sec.~1.9]{Damm2004}.
To test these predictions systematically, we consider game instances with $N=3$ players (and $N=2$ for comparison), $n=10$ and $m_i=2$, whose Gaussian data are scaled such that $\|B_i\|=\|A\|/2$, $\|C\|=s\|A\|$ and $\|D_i\|=s\|B_i\|$, with weights $Q_i$ and $R_{ii}$ of condition number $10$ and cross weights $R_{ij}$ that together have $40\,\%$ of the norm of the own weights.
For every game, $s_{\max}$ denotes the largest noise level at which it is mean-square stabilizable, computed by bisection on a linear matrix inequality \cite{aitrami_zhou_2000}.
Both schemes start from the same profile, the equilibrium of the preceding level if it is stabilizing and a stabilizing gain from this inequality otherwise, and a run is successful if all iterates are stabilizing and the residual of \eqref{eq:stationarity} drops below $10^{-8}$ within $1000$ iterations.

Fig.~\ref{fig:numerics}(a) follows the equilibrium of one game towards the stabilizability limit.
The rates of Propositions~\ref{prop:ibr_local} and~\ref{prop:ci_local} (solid) agree with the rates measured for the actual iteration maps (markers) to within $5\,\%$.
Left of the dotted line, the sensitivity $\mathcal A$ dominates, and both schemes contract at its rate, which is set by the costs.
Beyond it, the coupling $\mathcal J$ takes over, as described in Section~\ref{sec:regime}.
The rate of IBR then exceeds one, while the margin in the lower panel has hardly decreased.
At higher noise, it joins the dashed curve $\rho(\mathcal J_X)$, which depends only on the geometry of the noise channels and, by Corollary~\ref{cor:geometry}, can exceed one only for three or more players.
The rate of CI-PI, in contrast, hardly reacts to the regime change and tends to zero with the margin.
Close to the boundary, it decreases nearly in proportion to the margin, as Theorem~\ref{thm:rates}(ii) predicts.

Fig.~\ref{fig:numerics}(b) summarizes $1000$ such games at six noise levels and counts the runs in which at least one scheme found an equilibrium ($684$ to $974$ per level).
Whenever both schemes converged, they found the same equilibrium, and the hypotheses of Theorem~\ref{thm:rates} held at all computed equilibria.
CI-PI converges in almost all counted runs and needs its fewest iterations near the stabilizability limit, in line with its local rate.
An iteration of CI-PI took $2.4$\,ms and one of IBR about $12$\,ms.
The success rate of IBR mirrors the share of the equilibria that repel it, and both add up to almost $100\,\%$ at every level.
Indeed, the repelling equilibria account for $94$ to $99.5\,\%$ of the failures of IBR.
At $s=0.95\,s_{\max}$, the limit of Corollary~\ref{cor:geometry} identifies the repelling equilibria correctly in $99.9\,\%$ of the games.
The failure of IBR is therefore not a matter of the iteration budget or the initialization.
For $1000$ two-player games of the same family, in contrast, IBR is locally convergent at all $5785$ equilibria found across the six noise levels, as Corollary~\ref{cor:geometry}(a) predicts.
In these games, IBR fails almost only by exceeding the iteration limit.
The experiments confirm the regime change of Section~\ref{sec:regime}, in which the success rate of IBR drops from $80\,\%$ to below $1\,\%$, whereas CI-PI converges in at least $98.9\,\%$ of the counted runs at every noise level.

\section{Conclusion}\label{sec:conclusion}

This paper introduced policy iteration with coupled improvement.
CI-PI jointly solves the stationarity conditions of all players and thereby resolves the noise coupling.
Closed-form Fréchet derivatives revealed that the local dynamics of IBR remain driven by this coupling, even though exact best responses are computed.
The same holds for every scheme that improves the players against fixed strategies of the others.
As the noise intensity increases, a regime change occurs.
Beyond this change, the noise coupling governs local convergence, and CI-PI approaches superlinear convergence.
The convergence behavior of IBR, in contrast, is determined by the angles between the noise channels and may lead to local divergence when three or more players are involved.
Numerical experiments on three-player games confirm that this transition occurs at moderate noise levels. 
Moreover, near the stabilizability limit, IBR converged to fewer than $1\,\%$ of the computed equilibria, whereas CI-PI reached all of them.

\section*{Declaration of generative AI}
During the preparation of this work the authors used Claude (Anthropic) in order to
improve the language and the structure of the text.
After using this tool, the authors reviewed and edited the content as needed,
verified all mathematical statements and references, and take full responsibility
for the content of the published article.

\bibliographystyle{elsarticle-num}

\bibliography{paper/references}

\begin{thebibliography}{10}
\expandafter\ifx\csname url\endcsname\relax
  \def\url#1{\texttt{#1}}\fi
\expandafter\ifx\csname urlprefix\endcsname\relax\def\urlprefix{URL }\fi
\expandafter\ifx\csname href\endcsname\relax
  \def\href#1#2{#2} \def\path#1{#1}\fi

\bibitem{basar_olsder_1998}
T.~Ba{\c{s}}ar, G.~J. Olsder, Dynamic Noncooperative Game Theory, 2nd Edition,
  Vol.~23 of Classics in Applied Mathematics, SIAM, Philadelphia, PA, USA,
  1998.

\bibitem{hu_lauriere_2024}
R.~Hu, M.~Lauri{\`e}re, Recent developments in machine learning methods for
  stochastic control and games, Numerical Algebra, Control and Optimization
  14~(3) (2024) 435--525.

\bibitem{SAITO2019111}
T.~Saito, A.~Takahashi, Stochastic differential game in high frequency market,
  Automatica 104 (2019) 111--125.

\bibitem{10811852}
Z.~Liu, J.~Huang, Distributed {Nash} equilibrium seeking in aggregative games
  over jointly connected and weight-balanced networks, IEEE Transactions on
  Automatic Control 70~(5) (2025) 3486--3493.

\bibitem{10731297}
S.~Kille, P.~Leibold, P.~Karg, B.~Varga, S.~Hohmann,
  Human-variability-respecting optimal control for physical human-machine
  interaction, in: Proc. 33rd IEEE Int. Conf. on Robot and Human Interactive
  Communication (RO-MAN), 2024, pp. 1595--1602.

\bibitem{zhang_2026}
Z.~Zhang, K.~Handwerker, B.~Varga, A reinforcement learning framework for
  safety-aware haptic shared control in surgical robotics, arXiv:2610.15422
  (2026).

\bibitem{Todorov2005}
E.~Todorov, Stochastic optimal control and estimation methods adapted to the
  noise characteristics of the sensorimotor system, Neural Computation 17~(5)
  (2005) 1084--1108.

\bibitem{starr_ho_1969}
A.~W. Starr, Y.-C. Ho, Nonzero-sum differential games, Journal of Optimization
  Theory and Applications 3~(3) (1969) 184--206.

\bibitem{ZhuZhang2013}
H.~Zhu, C.~Zhang, Infinite time horizon nonzero-sum linear quadratic stochastic
  differential games with state and control-dependent noise, Journal of Control
  Theory and Applications 11~(4) (2013) 629--633.

\bibitem{basar_1987}
T.~Ba{\c{s}}ar, Relaxation techniques and asynchronous algorithms for on-line
  computation of non-cooperative equilibria, Journal of Economic Dynamics and
  Control 11~(4) (1987) 531--549.

\bibitem{HanHu2020}
J.~Han, R.~Hu, Deep fictitious play for finding {M}arkovian {N}ash equilibrium
  in multi-agent games, in: Proc. First Mathematical and Scientific Machine
  Learning Conference, Vol. 107 of Proceedings of Machine Learning Research,
  2020, pp. 221--245.

\bibitem{nortmann_2024}
B.~Nortmann, A.~Monti, M.~Sassano, T.~Mylvaganam, {Nash} equilibria for linear
  quadratic discrete-time dynamic games via iterative and data-driven
  algorithms, IEEE Transactions on Automatic Control 69~(10) (2024) 6561--6575.

\bibitem{chen_2026_tac}
Y.~Chen, C.~Chen, F.~L. Lewis, S.~Xie, Multiplayer cascaded policy iteration
  for {Nash} differential games, IEEE Transactions on Automatic Control 71~(2)
  (2026) 1038--1053.

\bibitem{guan_2024}
Y.~Guan, G.~Salizzoni, M.~Kamgarpour, T.~H. Summers, A policy iteration
  algorithm for {N}-player general-sum linear quadratic dynamic games, in:
  Proc. 63rd IEEE Conf. on Decision and Control (CDC), 2024, pp. 1725--1730.

\bibitem{thoemmes_2026}
F.~Th{\"o}mmes, L.~G{\"u}nther, K.~Handwerker, P.~Kr{\"u}ger, B.~Varga,
  S.~Hohmann, Policy gradient methods for continuous-time linear quadratic
  games, IFAC-PapersOnLine, in press (2026).

\bibitem{seqPI}
K.~Handwerker, F.~Th\"ommes, L.~G\"unther, B.~Varga, S.~Hohmann, Policy
  iteration for linear-quadratic stochastic differential games with state- and
  control-dependent noise, arXiv:2608.17940 (2026).

\bibitem{engwerda_2005}
J.~Engwerda, {LQ} Dynamic Optimization and Differential Games, Wiley, 2005.

\bibitem{basar_li_1989}
T.~Ba{\c{s}}ar, S.~Li, Distributed computation of {Nash} equilibria in
  linear-quadratic stochastic differential games, SIAM Journal on Control and
  Optimization 27~(3) (1989) 563--578.

\bibitem{srikant_basar_1992}
R.~Srikant, T.~Ba{\c{s}}ar, Sequential decomposition and policy iteration
  schemes for {M}-player games with partial weak coupling, Automatica 28~(1)
  (1992) 95--105.

\bibitem{Damm2004}
T.~Damm, Rational Matrix Equations in Stochastic Control, Vol. 297 of Lecture
  Notes in Control and Information Sciences, Springer, Berlin, Heidelberg,
  Germany, 2004.

\bibitem{Lucas}
L.~G{\"u}nther, K.~Handwerker, F.~Th{\"o}mmes, B.~Varga, S.~Hohmann,
  Infinite-horizon inverse linear-quadratic differential games with state- and
  control-dependent noise, arXiv:2608.17939v1 (2026).

\bibitem{aitrami_zhou_2000}
M.~{Ait Rami}, X.~Y. Zhou, Linear matrix inequalities, {Riccati} equations, and
  indefinite stochastic linear quadratic controls, IEEE Transactions on
  Automatic Control 45~(6) (2000) 1131--1143.

\bibitem{kleinman_1969}
D.~L. Kleinman, Optimal stationary control of linear systems with
  control-dependent noise, IEEE Transactions on Automatic Control 14~(6) (1969)
  673--677.

\bibitem{ortega_rheinboldt_1970}
J.~M. Ortega, W.~C. Rheinboldt, Iterative Solution of Nonlinear Equations in
  Several Variables, Academic Press, New York, USA, 1970.

\bibitem{HirschPughShub1977}
M.~W. Hirsch, C.~C. Pugh, M.~Shub, Invariant Manifolds, Vol. 583 of Lecture
  Notes in Mathematics, Springer, Berlin, Germany, 1977.

\bibitem{golub_vanloan_2013}
G.~H. Golub, C.~F. {Van Loan}, Matrix Computations, 4th Edition, Johns Hopkins
  University Press, Baltimore, MD, USA, 2013.

\bibitem{HornJohnson2012}
R.~A. Horn, C.~R. Johnson, Matrix Analysis, 2nd Edition, Cambridge University
  Press, 2012.

\bibitem{bauer_fike_1960}
F.~L. Bauer, C.~T. Fike, Norms and exclusion theorems, Numerische Mathematik 2
  (1960) 137--141.

\end{thebibliography}

\end{document}